\documentclass[3p,twocolumn]{elsarticle}

\usepackage[T1,T2A]{fontenc}
\usepackage[utf8]{inputenc} 
\usepackage[english]{babel}  

\usepackage{graphicx}    
\usepackage{subcaption}  
\usepackage{tikz}        
\usetikzlibrary{positioning} 

\usepackage{color} 

\usepackage{amsmath}
\usepackage{array} 
\usepackage{rotating} 
\usepackage{caption}

\graphicspath{{img/}}

\title{Impact of Electronic Energy Dissipation on Primary Radiation Damage Formation in Silicon}
\author{	
	N.~Korepanova$^{1*}$, R.~Nuñez-Palacio$^1$, A.E.~Sand$^{1*}$\\
	\footnotesize	$^1$\textit{Department of Applied Physics, Aalto University, P.O. Box 11100, 00076 Aalto, Finland}\\
	\footnotesize	${*}$ Corresponding authors. 
	\footnotesize	E-mail address: nadezda.korepanova@aalto.fi, andrea.sand@aalto.fi
}
\date{\today}

\makeatletter
\def\ps@pprintTitle{%
  \let\@oddhead\@empty
  \let\@evenhead\@empty
  \let\@oddfoot\@empty
  \let\@evenfoot\@oddfoot
}
\makeatother

\begin{document}

    \begin{abstract}		
        In this work, we investigate the role of ion-electron coupling in simulations of radiation damage formation in silicon using molecular dynamics simulations within a two-temperature model. We explore predictions of a threshold-free approach to the coupling that accounts for both the electronic stopping and electron-phonon coupling using a local electron density-based formalism. We compare two different coupling functions across a range of primary knock-on atom energies using two interatomic potentials.
	
        Our results demonstrate that the functional form of the ion-electron coupling plays a critical role in determining defect production efficiency, clustering, and recombination, and must therefore be carefully considered for accurate modeling of radiation damage formation. Furthermore, we find that the impact of the coupling in particular on the recombination of defects during the cooling phase of the cascade depends on the choice of interatomic potential, emphasizing the importance of physically grounded descriptions for both electronic effects and atom-atom interactions for reliable radiation damage predictions.
    \end{abstract}
	
	\maketitle	

	\section{Introduction}
	
	\begin{figure*}[h!]
		\includegraphics[scale=1]{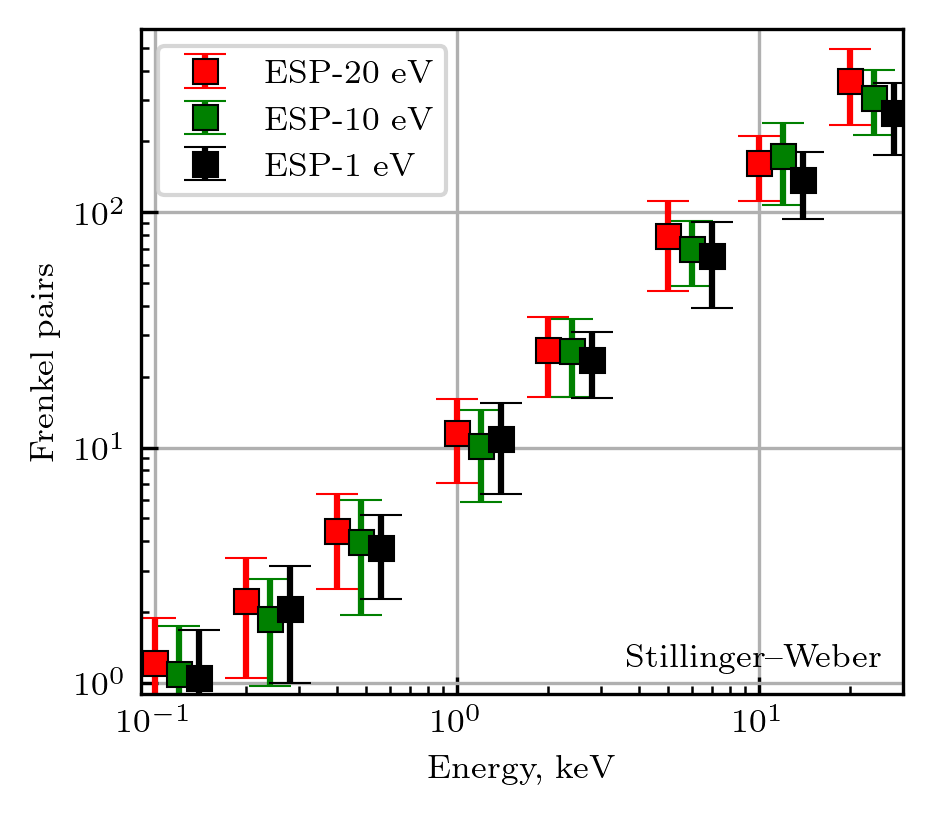}
		\includegraphics[scale=1]{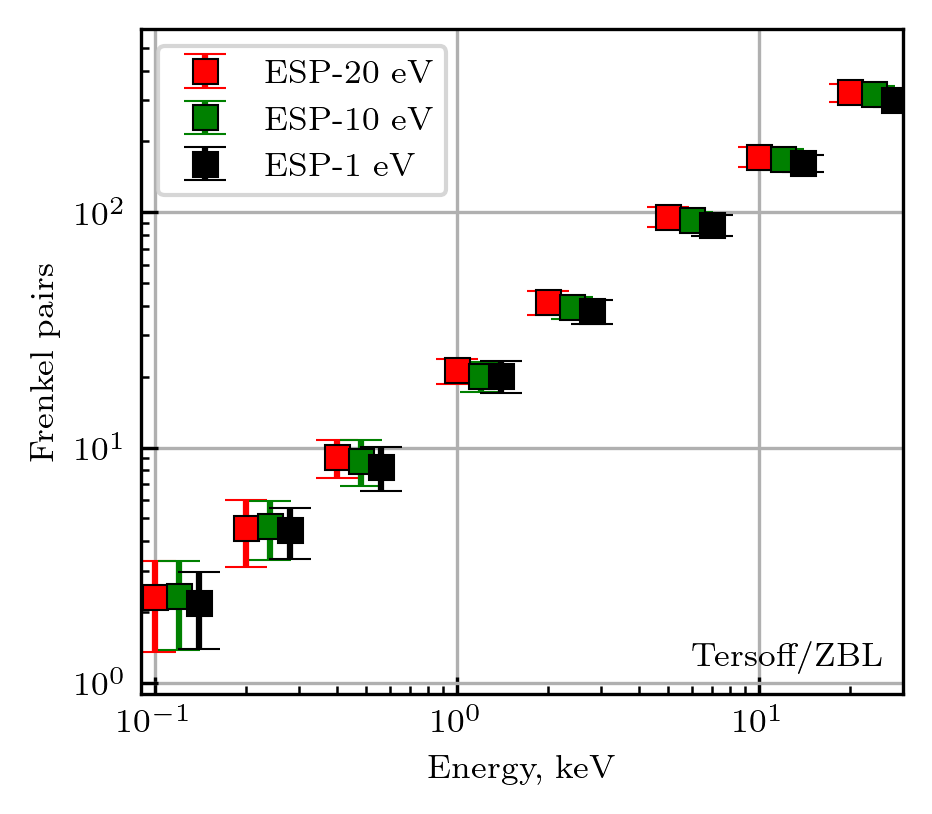}		
		\caption{Effect of kinetic energy cut-off on the number of Frenkel pairs produced in cascades as a function of initial PKA energy. $X$ in ESP-$X$ stands for the energy cut-off. The simulations were performed at the same PKA energies; however, the ESP-10\,eV and ESP-1\,eV data are shifted along the x-axis for clarity. }
		\label{fig:n-fp-esp}
	\end{figure*}
	
	Silicon is widely used in detectors and sensors operating in radiation-intensive environments such as space, nuclear reactors, and high-energy physics experiments. Radiation exposure progressively alters the structural and electronic properties of silicon, which compromises device functionality and can ultimately lead to failure. Understanding the mechanisms of radiation damage in silicon is therefore essential both for improving the reliability of the devices and for advancing the fundamental knowledge of damage formation and evolution in semiconductors.
	
	The radiation formation damage starts when an energetic particle elastically collides with lattice atoms, displacing them from their lattice sites and creating vacancies and interstitials, and their clusters. This displacement damage, formed in picoseconds, governs the long-term microstructural evolution and material degradation. 
	Because the creation of primary damage occurs within a fraction of a second, these processes are extremely challenging to capture experimentally. Consequently, molecular dynamics (MD) simulations have become a powerful tool for investigating primary damage formation, providing atomic-level insight into its mechanisms. However, MD typically relies on the assumption that electrons remain in their ground state during atomic motion, thereby neglecting the effects of the electronic system on damage production. 
	
    Classical MD simulations have significantly advanced the understanding of radiation damage in silicon, revealing the morphology of displacement cascades \cite{Caturla1995, Caturla1996, Averback1998}, defect clustering \cite{Stock1995, Otto2003}, and amorphization processes \cite{Nord2002}. These studies demonstrated that cascade evolution and defect production are strongly dependent on cascade conditions (energy, direction, and temperature) as well as on the chosen interatomic potential \cite{Otto2003, Nordlund1994, Nordlund1998}.  Common silicon potentials such as Stillinger-Weber and Tersoff yield different defect formation and recombination behavior, highlighting the sensitivity of radiation damage predictions to the underlying atomic interaction model.

    MD simulations have also clarified the microscopic mechanisms governing damage recovery and recrystallization in irradiated silicon, showing that highly disordered cascade regions can partially anneal during thermal equilibration on picosecond time scales \cite{Sayed1995, Denton1997}. At amorphous-crystalline interfaces, even low-energy recoil events can strongly influence recrystallization dynamics, reflecting a reduced displacement threshold in amorphous silicon \cite{Caturla1995_recrystal}.
	
    However, most of the early MD studies neglected explicit coupling between atomic motion and electronic excitations, effectively assuming that all kinetic energy deposited by the primary knock-on atom (PKA) remains within the lattice.

    In reality, as energetic atoms propagate through the material, they also transfer part of their energy to electrons, leading to additional slowing down of energetic atoms, i.e., electronic stopping ($S_e$).	
	Moreover, low-velocity atoms can exchange energy with electrons through electron-phonon coupling, which can affect the cooling rate during the thermal spike phase of displacement cascades by either transferring energy to the electronic subsystem or returning it to the ionic subsystem. The role of energy transfer to electrons in defect production and evolution is complex and not well understood.
	
	In earlier studies, single crystal silicon was considered to be insensitive to electronic excitations; neither recrystallization of amorphous domains \cite{Chavan1995_no_recrystalization} nor latent track formation was observed experimentally. Experiments with swift heavy ions (SHI) \cite{Toulemonde1989, Levalois1992_no_recrystallization} showed that damage creation due to electronic excitation is negligible up to $S_e$ of $\approx$14 keV/nm.
	Later, the formation of ion tracks in Si was observed for energetic cluster ions, which produce higher $S_e$ values compared to monoatomic ions, 30 keV/nm \cite{Canut1998_exp_si_sensitivity_to_electr_exitation} and 46 keV/nm \cite{Dunlop1998}. 
	In more recent work, however, ion track formation in Si has also been observed at much lower cluster energies ($S_e$ of $\approx$5-18.2 keV/nm) \cite{Amekura2021}.	
	This finding was attributed to synergistic effects between nuclear and electronic stopping, since no ion track formation was observed under irradiation with 200 MeV Xe ions ($S_e$ = 14 keV/nm) in the same study. More recently, defect recovery and recrystallization induced by local electronic excitation have been observed in pre-damaged Si exposed to monoatomic irradiation \cite{Mihai2023_exp_ion_induced_recrystalization, Mihai2019}, highlighting the significant role of electronic excitation processes in annealing.
	
	To account for electronic effects in MD simulations, several models have been developed that extend the standard MD framework. The simplest approach introduces a friction force acting on energetic atoms; however, this requires a kinetic energy cut-off ($T_c$) below which the friction is not applied, to avoid unphysical damping of thermal atomic motion. Various values of $T_c$ have been used in the literature, including 1\,eV~\cite{Page2009, Bjorkas2009}, 10\,eV~\cite{Sand2013, Sand2015, Kim1988}, or twice the material-specific cohesive energy~\cite{Zarkadoula2016}. Although the choice of the cut-off value leads to a quantitative effect on predictions \cite{Sand2013, Sand2015}, there is no clear consensus on an appropriate value. 
	Fig.\ref{fig:n-fp-esp} shows the effect of the cut-off value on the prediction of radiation damage in silicon.
		
	Although the above approach accounts for electronic stopping by removing kinetic energy from atoms, it does not explicitly model energy transfer to the electronic subsystem.
	The exchange of energy between the atomic and electronic subsystems can be implemented via a two-temperature model (TTM), such as that developed by Duffy and Rutherford \cite{Duffy2006} for radiation damage simulations in MD, loosely based on earlier work by Caro and Victoria \cite{Caro1989}. 	
	This model accounts for electronic stopping through a frictional force and for electron-phonon (e-ph) coupling through a stochastic term that enables energy transfer between the atomic lattice and the electronic subsystem. A cut-off velocity is introduced, above which both electronic stopping and electron-ion interactions are included, while below this threshold only the electron-phonon coupling term is applied.
	Later in \cite{Zarkadoula2017}, a time threshold for activating the electron-phonon coupling was also introduced as an additional parameter, motivated by the longer timescale of the e-ph coupling compared to the early stages of cascade dynamics. Parametric studies \cite{Jarrin2020, Zhou2020, Rojano2024} show that the choice of time threshold and cut-off velocity significantly affects the predictions of cascade simulations, but physically justified values for these parameters remain lacking.

    However, in their original work, Caro and Victoria \cite{Caro1989} postulated that the same physical mechanisms govern both electronic stopping and electron-phonon coupling, and suggested that energy losses could be  captured by a single coupling term dependent on the local electron density. Consequently, they introduced a function that incorporates the local electronic density within the simulation domain, leading to a threshold-free dynamic coupling that depends directly on the local environment of each ion.

	The electron-density-dependent coupling scheme within a two-temperature framework was subsequently implemented by Tamm et al. \cite{Tamm2018, Tamm2019} as a user plug-in (USER-EPH)  \cite{eph-web} for the LAMMPS code \cite{lammps,lammps-web}. Within this approach, the dependence of energy losses on the local electronic density is determined from first-principles real-time (rt) time-dependent density functional theory (TDDFT) calculations, from which the coupling functions are then fitted and incorporated into the model. In the following, this approach is referred to as the unified two-temperature model (UTTM), to differentiate it from the 'TTM' model by Duffy and Rutherford.
	
	For silicon, several coupling functions were proposed in \cite{Jarrin2021}, demonstrating that the UTTM can reproduce TDDFT-predicted electronic stopping within MD simulations. To this end, simple parameterizations, including constant and quadratic functional forms, were used. 
	The authors also compared cascade simulations using different UTTM and TTM parameterizations, showing that UTTM leads to significantly different defect production compared to TTM. However, the impact of electronic effects on defect production and evolution was not examined in detail.
	
    In this work, we investigate the effect of electronic energy losses on cascade dynamics and the sensitivity of predictions of the formation of primary radiation damage on the ion-electron coupling model in diamond-structured silicon. To this end, we compare two different coupling functions within the UTTM framework: the quadratic function of Ref.~\cite{Jarrin2021}, and a recent coupling function developed by us to capture energy losses for all trajectories with high fidelity, including close collisions \textbf{\cite{Nunez_data, Nunez_paper}}. The latter coupling was created through a fitting procedure that identifies four key regions of the electronic density and will thus be referred to in the following as the 'four-density' coupling.
	
	The paper is organized as follows. Section~\ref{sec:method} describes the simulation details and analysis methodology. Section~\ref{sec:method} presents and discusses the results. Section~\ref{sec:conclusion} summarizes the main findings.

	\section{Simulation details and analysis}
    \label{sec:method}
	
	\begin{table*}[ht]
		\centering
		\caption{PKA energies and simulation cell characteristics for displacement cascades in silicon}
		\label{table:box-size}
		\begin{tabular}{lll}
			\hline
			PKA energy, keV & Simulation box size, $a_0$ & Number of atoms \\
			\hline
			0.1  & 8 x 8 x 8       &      4 096 \\
			0.2  & 10 x 10 x 10    &      8 000 \\
			0.4  & 14 x 14 x 14    &     21 952 \\
			1.0  & 18 x 18 x 18    &     46 656 \\
			2.0  & 22 x 22 x 22    &     85 184 \\
			5.0  & 50 x 50 x 50    &  1 000 000 \\
			10.0 & 100 x 100 x 100 &  8 000 000 \\
			20.0 & 120 x 120 x 120 & 13 824 000 \\
			\hline
		\end{tabular}
	\end{table*}

	We performed collision cascade simulations with the LAMMPS code \cite{lammps,lammps-web} together with the USER-EPH plugin \cite{eph-web} for PKA energies of 0.1, 0.2, 0.4, 1.0, 2.0, 5.0, 10.0, and 20.0~keV. 	
	We employed two different semi-empirical interatomic potentials that are frequently applied in the field of radiation damage: the Stillinger-Weber (SW) potential \cite{sw} and the Tersoff potential smoothly joined to the universal Ziegler-Biersack-Littmark repulsive potential at short interatomic distances (T3/ZBL) \cite{tersoff-zbl}.
	
	The cascades were simulated in a cubic system with periodic boundaries in all directions. The silicon atoms were arranged in a diamond cubic lattice. The simulation cell was initially equilibrated at 300 K and 0 kbar using the Nose/Hoover temperature thermostat and the Nose/Hoover pressure barostat \cite{Nose1984, Hoover1985, Martyna1994}.
	A Langevin thermostat was applied to the border atoms during cascade simulations to enable the dissipation of excess heat introduced by the PKA into the bulk of the material. The cell boundaries were monitored to ensure that no energetic atoms crossed the periodic boundaries. This guaranteed that the cascades remained within the cell and did not self-interact. The sizes of the simulation cells for each PKA energy are listed in Table \ref{table:box-size}.
	
	Each cascade was initiated by assigning kinetic energy to a PKA atom along a randomly selected direction. The PKA was chosen so that the velocity vector pointed toward the center of the simulation box. The equations of motion were integrated using adaptive time steps, with the maximum allowed time step of 0.0005\,ps. The system was allowed to evolve until the number of defects reached a steady value, indicating the end of the recombination phase of the cascade. 	
	
	The electronic system was constructed to be three times larger than the atomic system in each direction. It was discretized into voxels with a side length of $\approx$25,\AA, providing spatial resolution for the electronic temperature field. The initial temperature of the electronic system was set to 300\,K to match that of the atomic system. The electronic heat capacity ($C_e$) and the electronic thermal conductivity ($\kappa_e$) were set to $5\times10^{-6}$\,eV/K/\AA$^3$ and $5\times10^{-3}$\,eV/K/\AA/ps, respectively, corresponding to low electronic temperatures \cite{Jarrin2021}. To investigate the impact of energy feedback from the electronic system to the atomic system, we performed simulations with a fixed electronic temperature, in which energy is mainly transferred only from the atomic to the electronic subsystem, and with a variable electronic temperature, which allows the electronic system to heat up leading to stronger bidirectional energy exchange between the two subsystems. 
    
    We employed two different coupling functions: the quadratic form reported in \cite{Jarrin2021} and the recently developed four-density coupling function \cite{Nunez_data, Nunez_paper}. These implementations of the UTTM coupling have different focus, and hence represent different levels of fidelity with respect to rt-TDDFT predictions on which they are fitted.

	To fit the quadratic coupling function, the authors in \cite{Jarrin2021} considered ten representative trajectories in Si: central channels along the $\langle 001 \rangle$, $\langle 110 \rangle$, and $\langle 111 \rangle$ directions; off-center variants of the $\langle 001 \rangle$ and $\langle 110 \rangle$ channels; and four incommensurate directions, including one containing a vacancy. The electron density $\rho(r)$ of an atom in vacuum was used as a baseline for the parameterization of the coupling function. The coupling function was assumed to have a quadratic dependence on the electron density, given by $(a\rho + b\rho^2 )/(1 + e^{c(\rho-\rho_f)})$. The parameters were optimized to minimize the mean absolute error between the UTTM-MD energy-loss-versus-distance curves and the corresponding TDDFT results across all sampled trajectories.	
	The resulting best-fit ``quadratic'' form effectively reduces to a linear-with-saturation expression, as the optimization drives the quadratic coefficient $b$ to zero: $(0.041\rho + 0.0\rho^2)/(1 + e^{10(\rho - 0.3)})$ eV$\cdot$ps/\r{A}$^2$. 
	Despite this, the term ``quadratic'' is retained to reflect the original functional form and is used in this work for consistency.

	To construct the four-density coupling function, the electronic stopping powers of eight representative trajectories were considered: the crystallographic channels $\langle 110\rangle$, $\langle 001\rangle$, $\langle 112\rangle$, and $\langle 111\rangle$; the half-centered $\langle 111\rangle$ channel; the off-centered $\langle 110\rangle$ channel; and two incommensurate paths, one of which involves a collision-like approach to a lattice atom. The density $\rho(r)$ was constructed to align with the electron density in the diamond lattice. The resulting density dependence of the coupling function has four electron-density regions corresponding to distinct local environments in the diamond structure: (i) regions with three or more equally spaced neighbors, or out-of-bond regions; (ii) regions with two equally spaced neighbors, corresponding to bond regions;  (iii) regions with a single dominant neighbor, corresponding to the core regions; and (iv) regions approaching the nucleus of a host atom or deep-core regions, where the dissipative process is strongest. Further details on the four-density coupling function can be found in \cite{Nunez_paper}.
	
	Additional simulations implementing electronic stopping as a friction force were carried out, and are used here primarily as a reference. In this approach, the stopping power is obtained from the SRIM software \cite{srim-web} and implemented in the simulations as a velocity-dependent damping term \cite{Nordlund1995, Nordlund1998, lammps_elstop_fix}, and no two-temperature framework is invoked. We consider kinetic energy cut-offs of 1, 10, and 20\,eV, denoted as ESP-1eV, ESP-10eV, and ESP-20eV, respectively.	
	The cumulative energy loss to electrons is evaluated in LAMMPS for both the friction-based electronic stopping simulations and the UTTM simulations.

    \begin{figure*}[h!]
		\centering
		\includegraphics[scale=1]{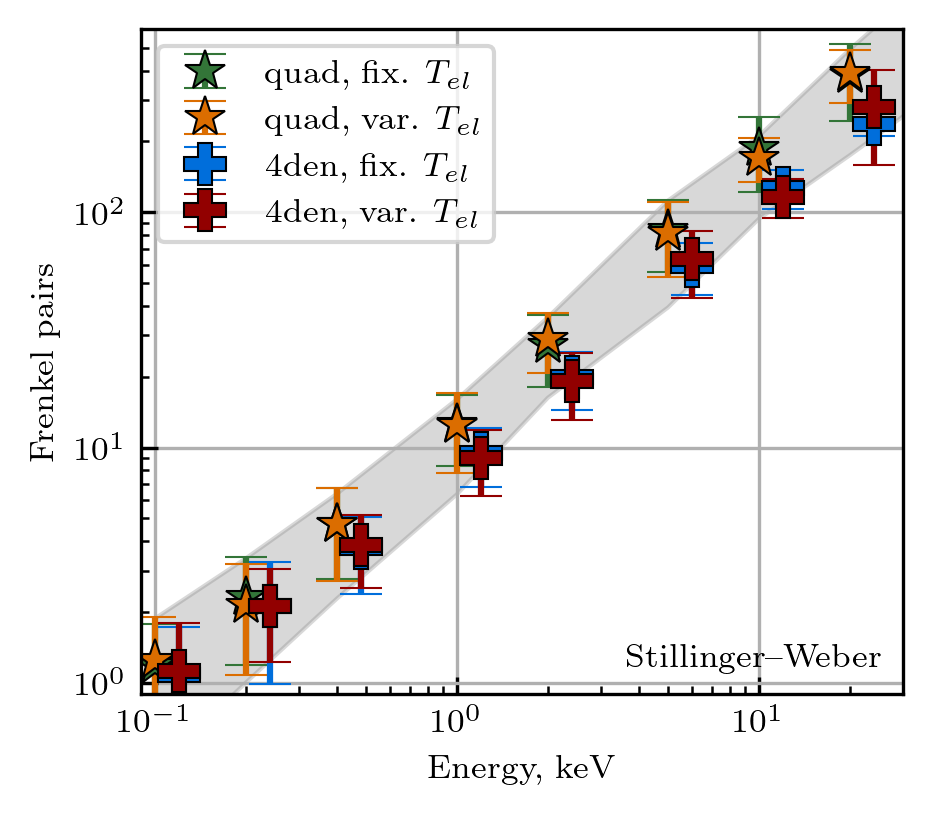}
		\includegraphics[scale=1]{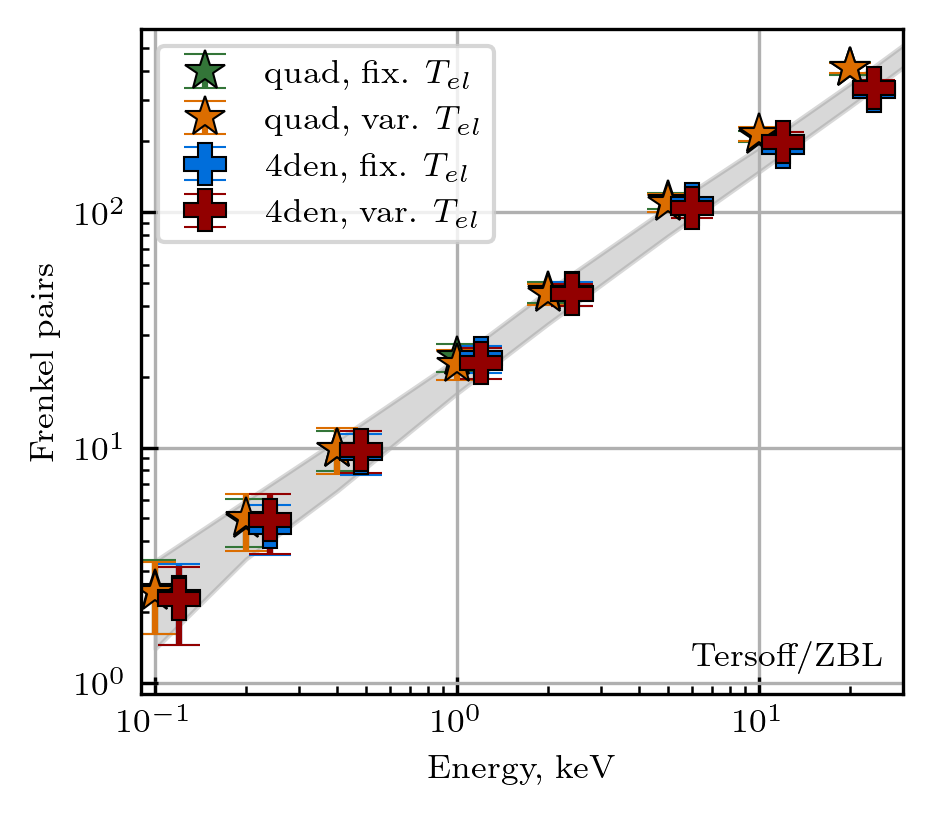}
		\caption{Effect of the coupling function on the number of Frenkel pairs produced in cascades as a function of initial PKA energy. The simulations were performed at the same PKA energies; however, the four-density coupling function data are shifted along the x-axis for clarity. The gray area represents the range of predictions from simulations in which electronic losses are modeled as a friction force. }
		\label{fig:n-fp}
	\end{figure*}

    \begin{figure*}[h!]
		\includegraphics[scale=1]{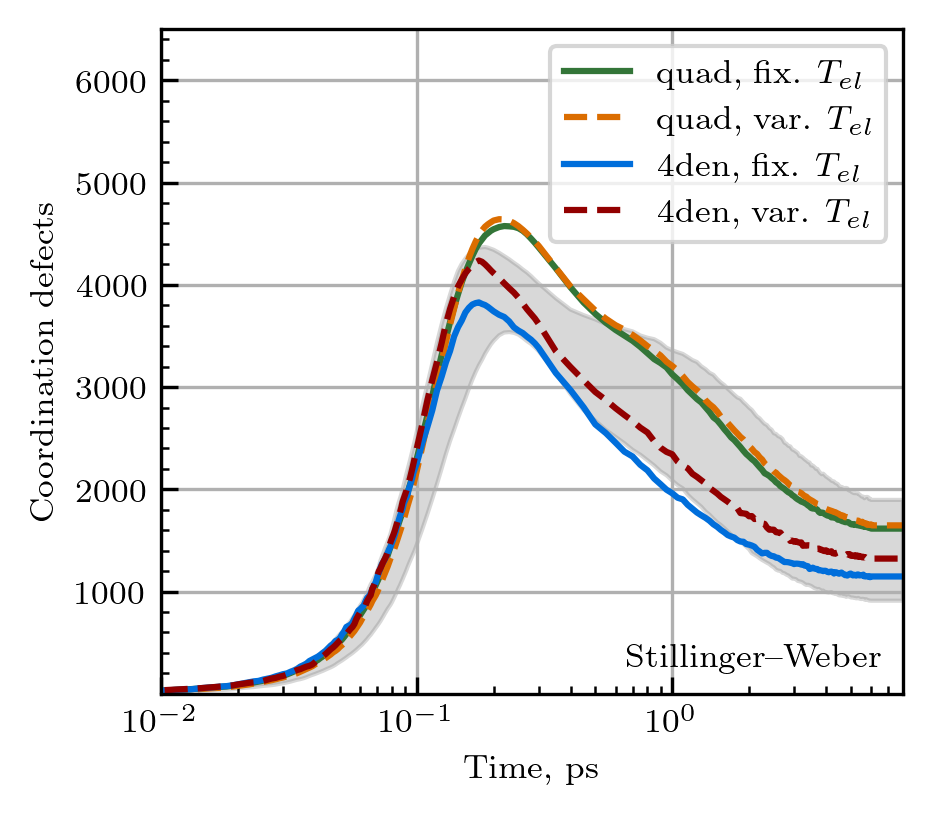}
		\includegraphics[scale=1]{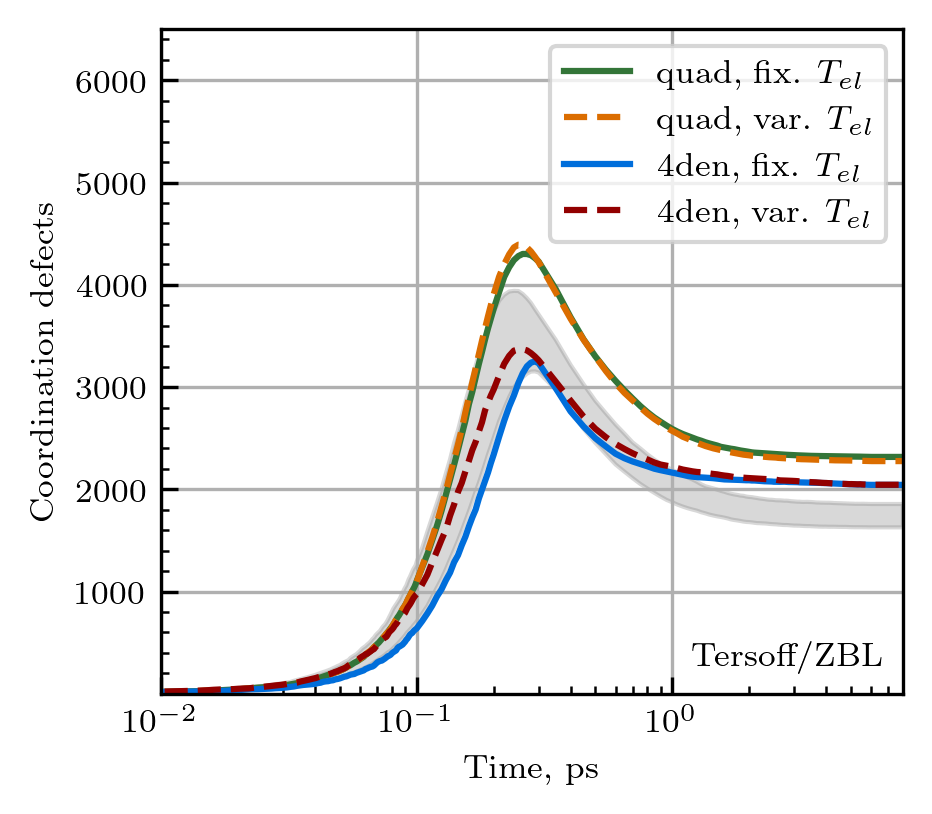}
		\caption{Instantaneous number of coordination defects. The gray area represents the range of predictions from simulations in which electronic losses are modeled as a friction force. Results are averaged among all cascades with PKA energy of 20\,keV.}
		\label{fig:coord-def-time-evolution}
	\end{figure*}

	Residual defects were identified using the Wigner-Seitz (WS) analysis method implemented in OVITO \cite{ovito}. Coordinated defects were defined as atoms having either more or fewer than four neighboring atoms within the cutoff radius. In this study, we used a cutoff radius of r~=~2.85\,\AA~\cite{Hensel1998}.		

    Cluster analysis was carried out by defining a cluster as a set of neighboring Wigner–Seitz defects located within a distance up to 
	$r_{cl}=2a_0 \approx 10.8$\,\AA~\cite{Nordlund1998}.
	Many clusters contain both vacancies and interstitials; therefore, the cluster size is defined as the total number of defects it contains, including both vacancies and interstitials.
	The spatial extent of a cluster was determined using principal component analysis (PCA). The defect coordinates were projected onto the first principal component, and the length was defined as the range of projections along this axis.

    \begin{figure}[t]
		\includegraphics[scale=1]{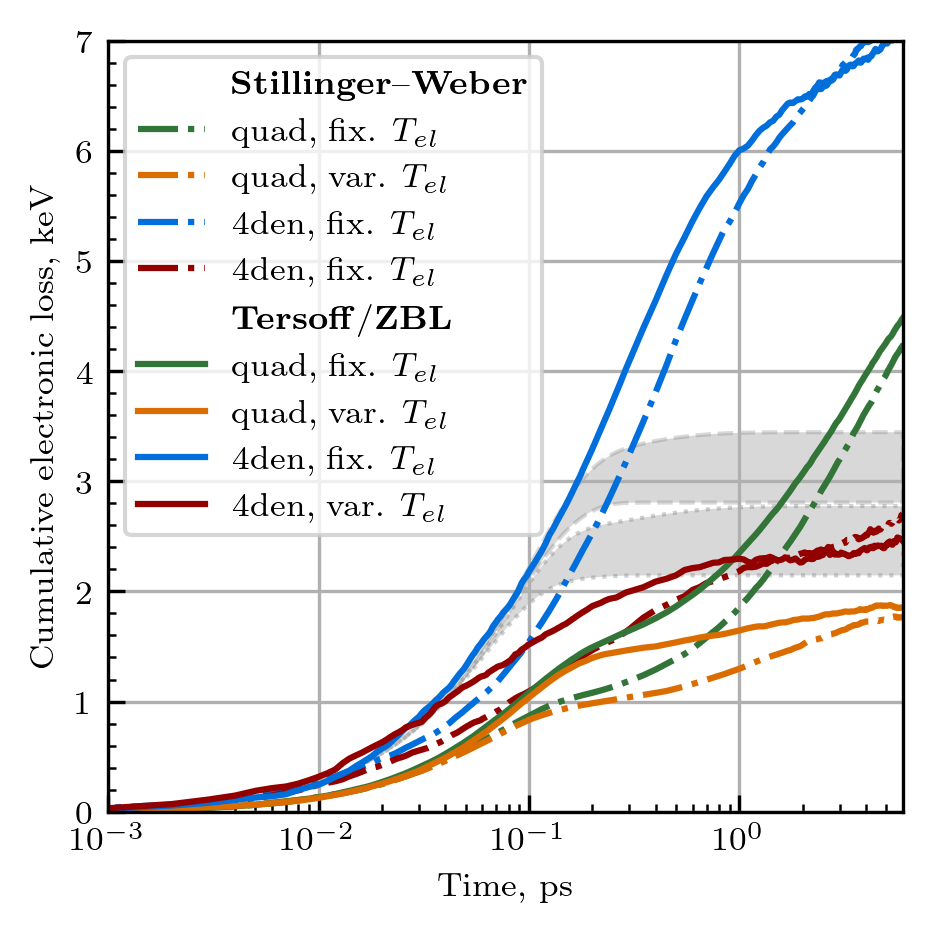}
		\caption{Cumulative energy loss to electrons obtained from MD simulations.  
			The gray area represents the range of averaged energy loss from simulations in which electronic losses are modeled as a friction force (for ESP-1\,eV and ESP-20\,eV). With a lower filled area for the Stillinger-Weber potential.
			Results are averaged for cascades with PKA energy of 20\,keV.			
			}
		\label{fig:cumulative-elloss}
	\end{figure}

    \begin{figure}[h]
		\centering
		\includegraphics[scale=1]{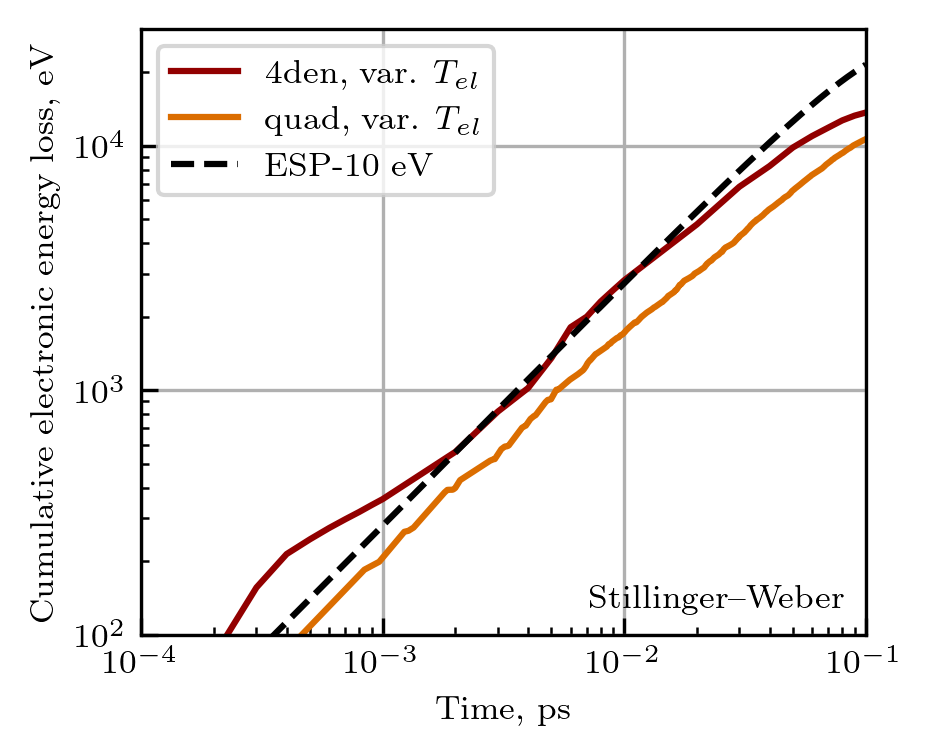}
		\caption{Cumulative energy loss to electrons obtained from MD simulations. $10$ in ESP-10\,eV stands for the energy cut-off for $S_e$. Results are averaged for cascades with PKA energy of 100\,keV.}
		\label{fig:elloss-100kev}
	\end{figure}

	\section{Results}
    \label{sec:results}
	
	\subsection{Damage production}

    \begin{figure}[h!]
		\centering
        \includegraphics[scale=1]{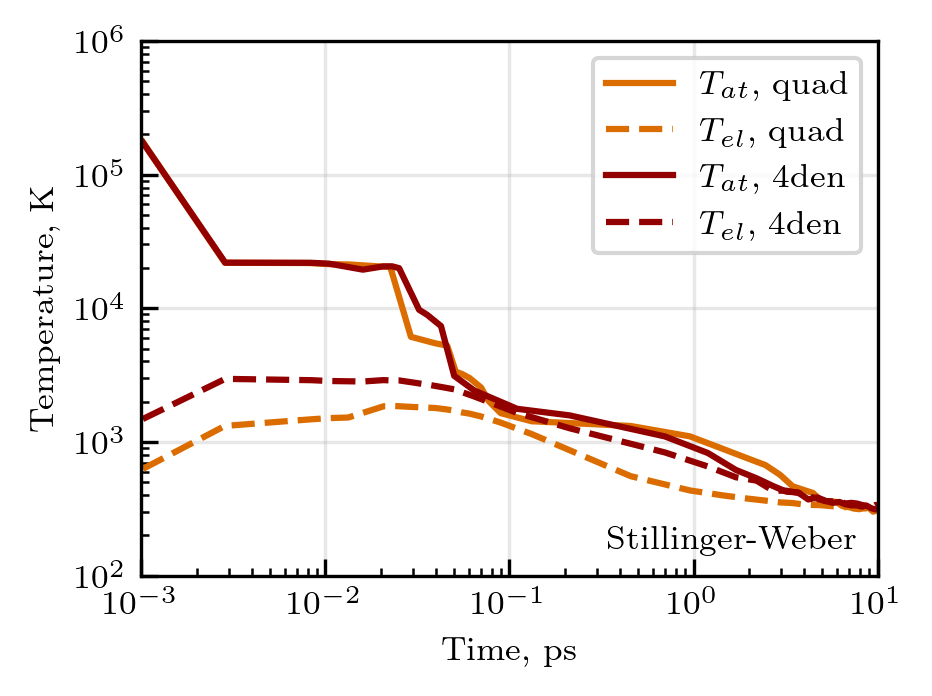}
		\caption{Atomic and electronic temperature of voxel where cascade was initiated at different times. For four-density and quadratic coupling function. Results are shown for a single 20\,keV cascade with identical initial conditions for both coupling functions. }
		\label{fig:atom-electron-temp}
	\end{figure}
    
	The average number of surviving Frenkel pairs (FPs) as a function of PKA energy is shown in Figs.~\ref{fig:n-fp}. MD simulations using the UTTM models with a quadratic coupling function predict a higher number of FPs than simulations with the four-density coupling function. At the same time, the difference between using a varying or a fixed electronic temperature remains small. This indicates that for the chosen conditions ($C_e$, $\kappa_e$, temperature), the functional form of the coupling and the resulting energy losses to the electronic system have a stronger influence on defect production than the subsequent feedback from a heated electronic system to the lattice. The cascade evolution is shown in Fig.~\ref{fig:coord-def-time-evolution} and further highlights these differences. The quadratic coupling function leads to a higher peak number of coordination defects, indicating that more energy remains in the atomic subsystem during the ballistic and thermal spike phases.  
	In contrast, the four-density coupling function reduces the number of peak defects. The number of surviving coordination defects follows the same overall trend.
			
	The difference in defect production correlates with the different electronic energy losses. Fig.~\ref{fig:cumulative-elloss} shows the cumulative energy losses to the electronic subsystem predicted by UTTM-MD employing different coupling functions. The four-density coupling function predicts significantly higher energy transfer to the electronic subsystem during the ballistic phase. We find that the energy losses during the ballistic phase are critical for defect production, with enhanced energy dissipation reducing the amount of energy available for the cascade to propagate through the lattice and create damage regions. In the ballistic regime, the four-density coupling function predictions closely follow energy losses obtained using SRIM stopping powers, especially for the T3/ZBL potential, and in the fixed electronic-temperature case, simulations using the four-density coupling function continue to follow SRIM for a longer time. However, once ESP energy losses have levelled off, those predicted by the four-density coupling function continue to increase, since the UTTM formulation does not impose an effective limit on energy transfer to the electronic subsystem. 

	Comparison of the electronic stopping power predicted by MD simulations using the four-density and quadratic coupling functions shows that the four-density coupling function is in better agreement with the TDDFT predictions~\cite{Nunez_paper}. This suggests that the four-density formulation provides a more accurate description of energy transfer between moving atoms and the electronic subsystem for the complex trajectories involved in the cascade event.	
	An additional indication is provided by the energy losses during the ballistic phase of displacement cascades, when atoms move with high kinetic energies and energy dissipation is dominated by electronic stopping along random trajectories traveling largely through the pristine lattice. Since SRIM stopping powers are based on extensive experimental datasets, mainly obtained from energetic projectiles traveling along straight paths through an intact material, they provide a useful reference for electronic energy losses in this regime. Consequently, a model that deviates significantly from SRIM may introduce unphysical electronic energy loss rates. 
	Figure~\ref{fig:elloss-100kev} shows the electronic energy losses predicted by different models during the ballistic stage of a 100\,keV cascade. This energy was chosen because experimental stopping-power data are available in this energy range \cite{Lohmann2020} and SRIM is known to reproduce the measurements with good accuracy \cite{Lohmann2020}.
	The good agreement of the four-density coupling function with SRIM stopping powers therefore further supports its physical consistency in the ballistic regime.

	The more efficient energy loss predicted with the four-density coupling arises from the stronger coupling between atomic and electronic subsystems compared to the quadratic model. Figure~\ref{fig:atom-electron-temp} shows the time evolution of the atomic and electronic temperatures of the voxel where the cascade was initiated. For the same initial atomic temperature, the four-density coupling function produces a higher rise in electronic temperature, followed by a more efficient equilibration between atomic and electronic subsystems. In contrast, the quadratic coupling leads to weaker energy transfer and a larger temperature mismatch, allowing more energy to remain in the lattice.

	Although both coupling functions are fitted to rt-TDDFT data, they differ significantly in their transferability to cascade conditions. Under cascade conditions, the choice of functional form and the trajectories used for fitting become important. In particular, the four-density coupling function was fitted using trajectories with close collision approaches, which improves its transferability to cascade conditions.
	
	\subsection{Clustering}
	
	\begin{figure*}[h!]
		\centering
		\includegraphics[scale=1]{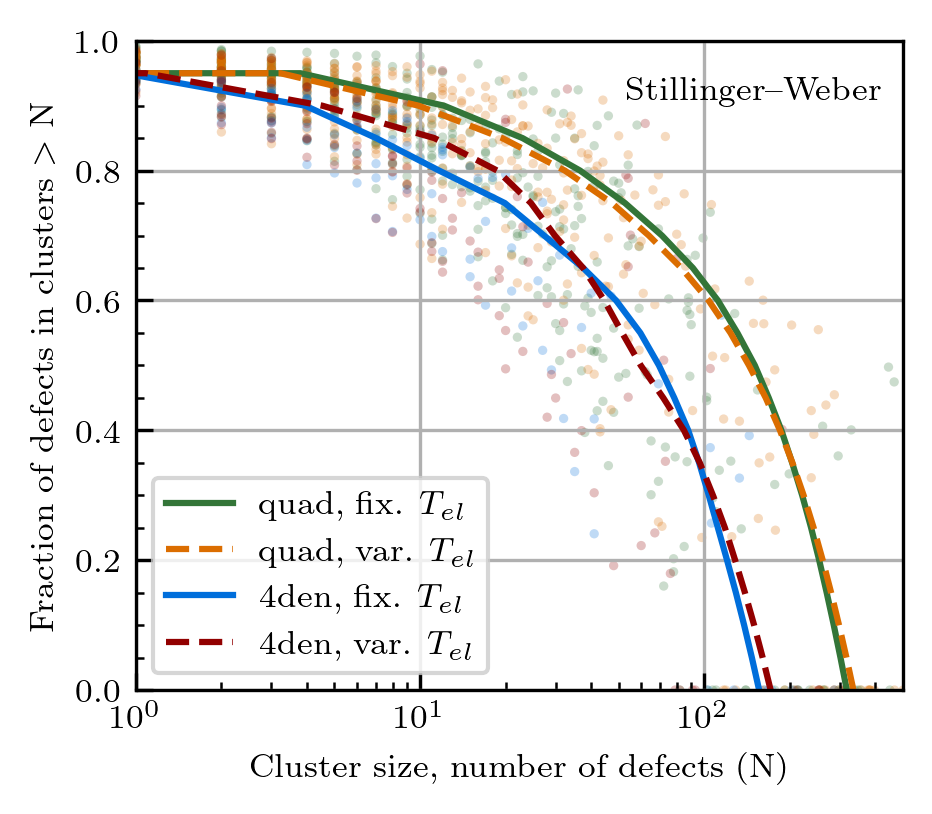}
		\includegraphics[scale=1]{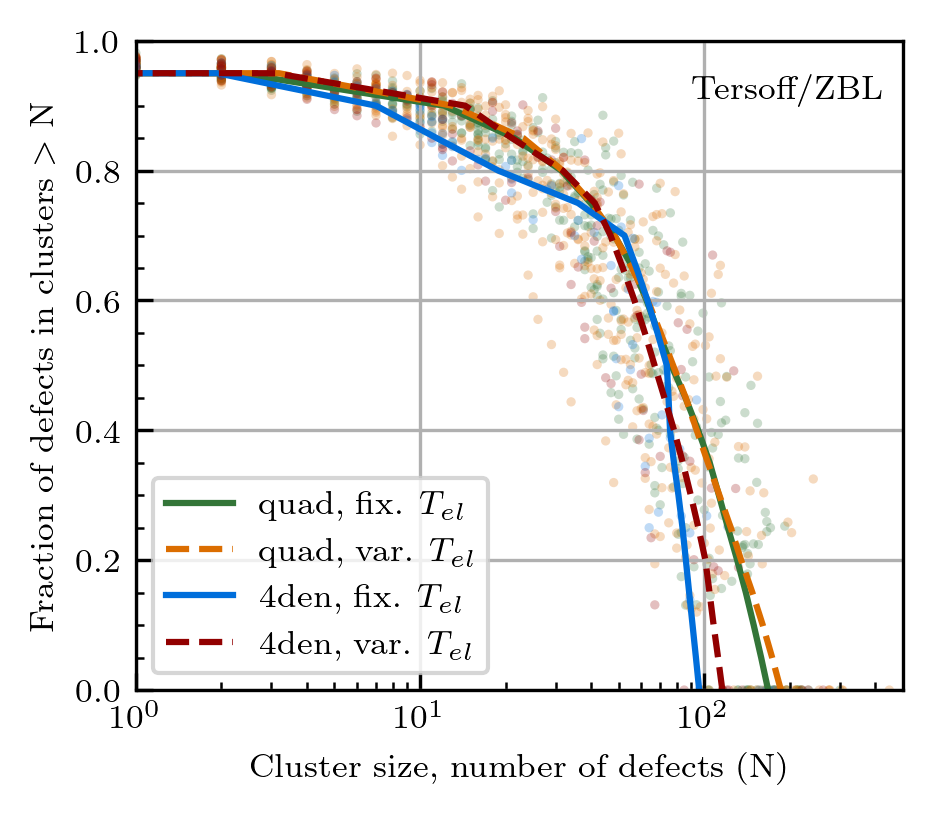}
		\caption{The fraction of defects in clusters larger than N. Results are averaged for cascades with PKA energy of 20\,keV.}
		\label{fig:clustering}
	\end{figure*}
	
	\begin{figure}[t]
		\includegraphics[scale=1]{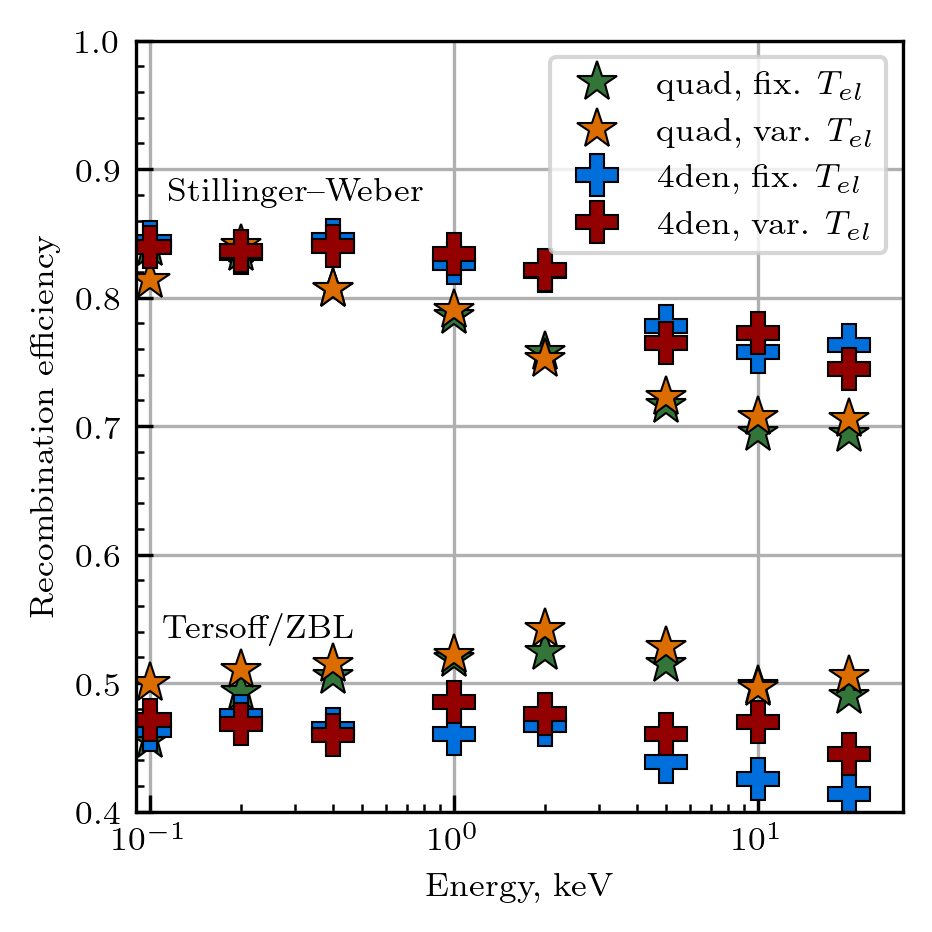}
		\caption{The recombination efficiency in the cascade.}
		\label{fig:recysltal}
	\end{figure}
	
	\begin{figure*}[h!]
		\includegraphics[scale=1]{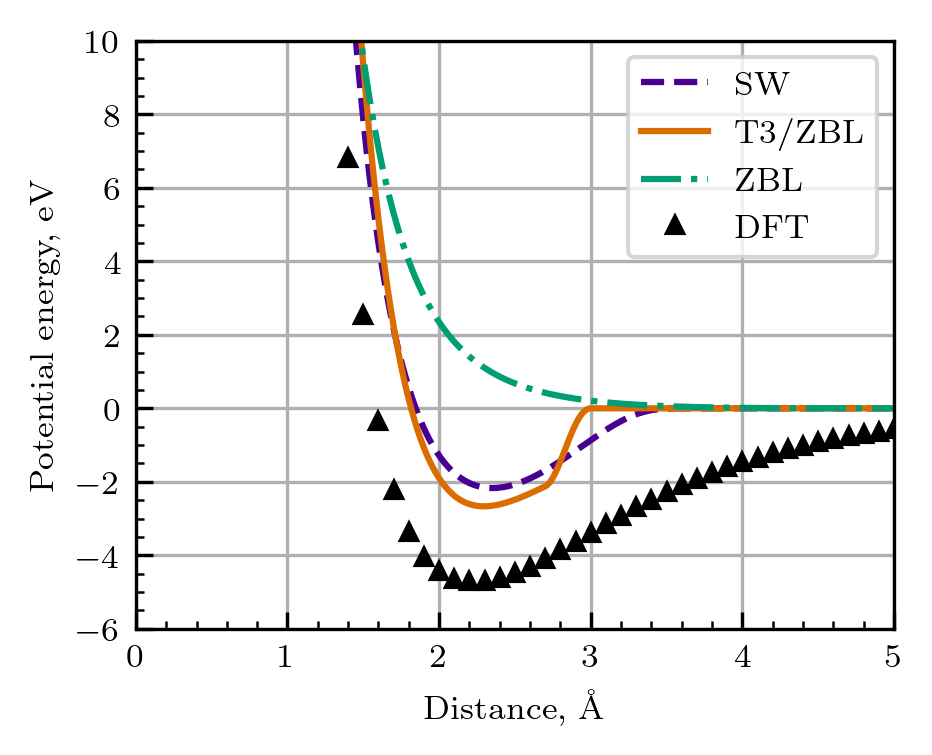}
		\includegraphics[scale=1]{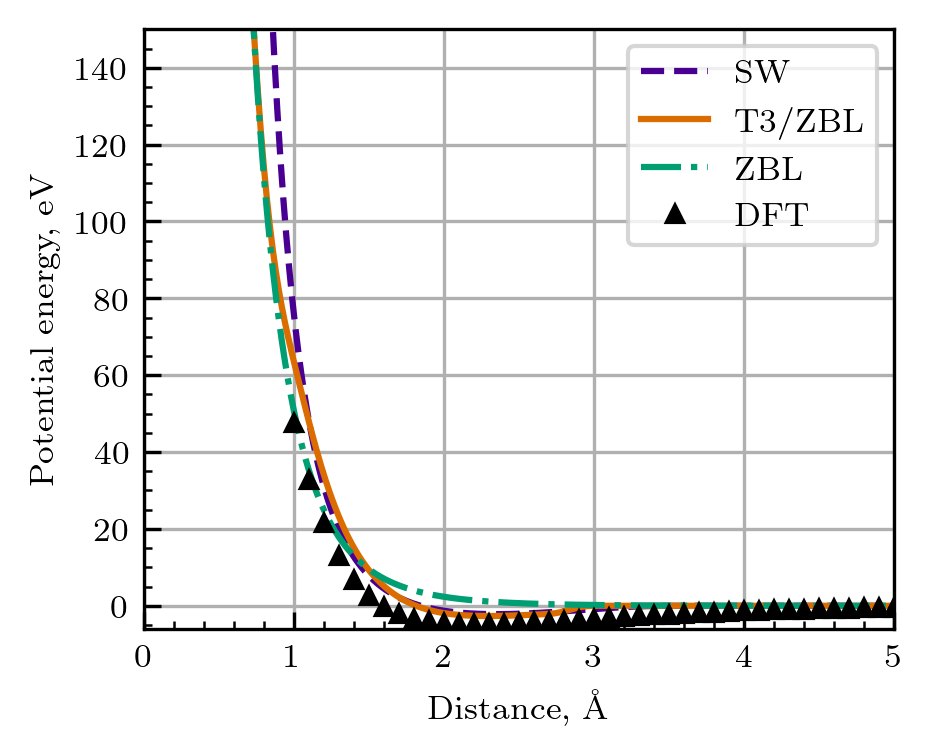}
		\caption{Potential energy of a dimer. The DFT data is taken from  \cite{Saha2026}.
		}
		\label{fig:dimer-poteng}
	\end{figure*}

	\begin{figure*}[h!]
		\centering
		\includegraphics[scale=1]{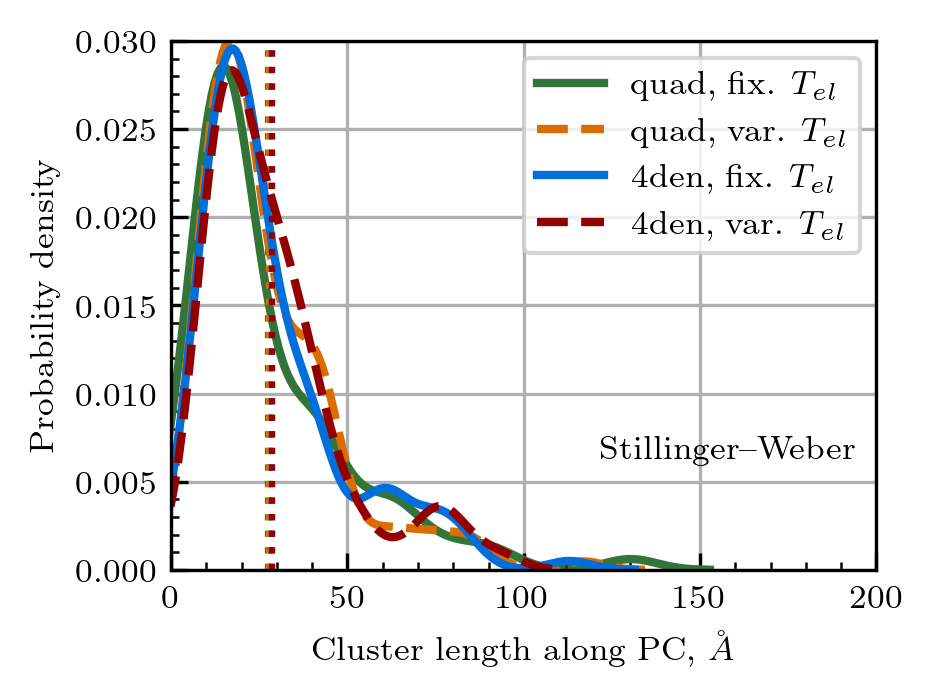}
		\includegraphics[scale=1]{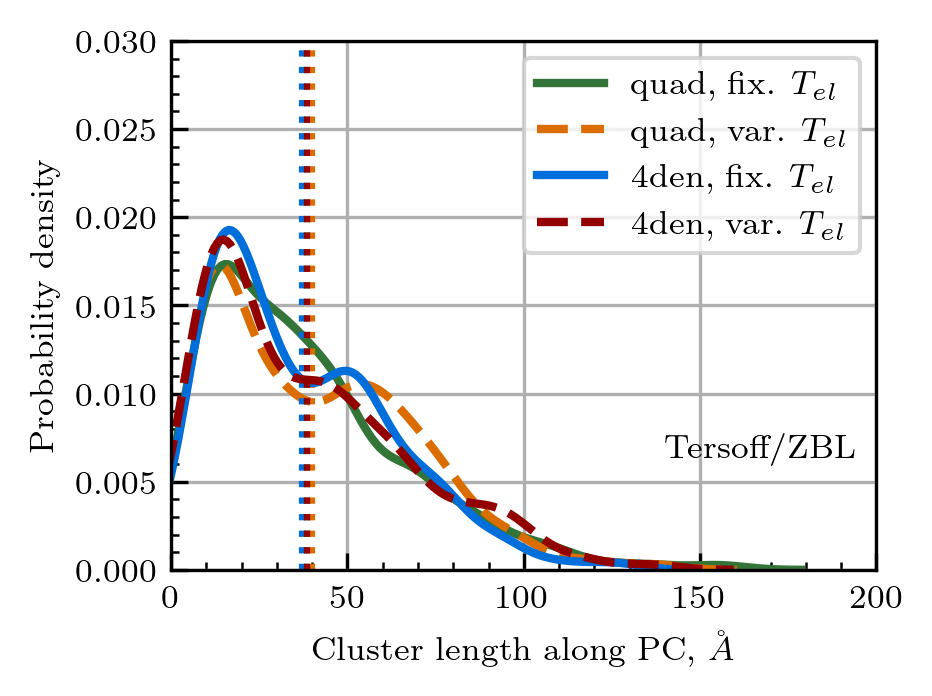}
		\includegraphics[scale=1]{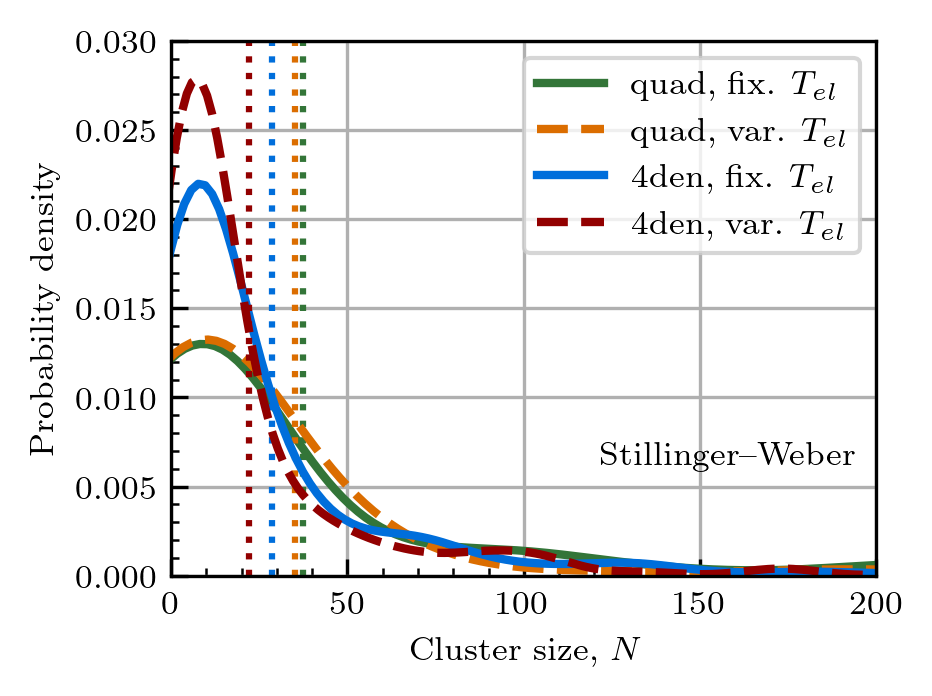}
		\includegraphics[scale=1]{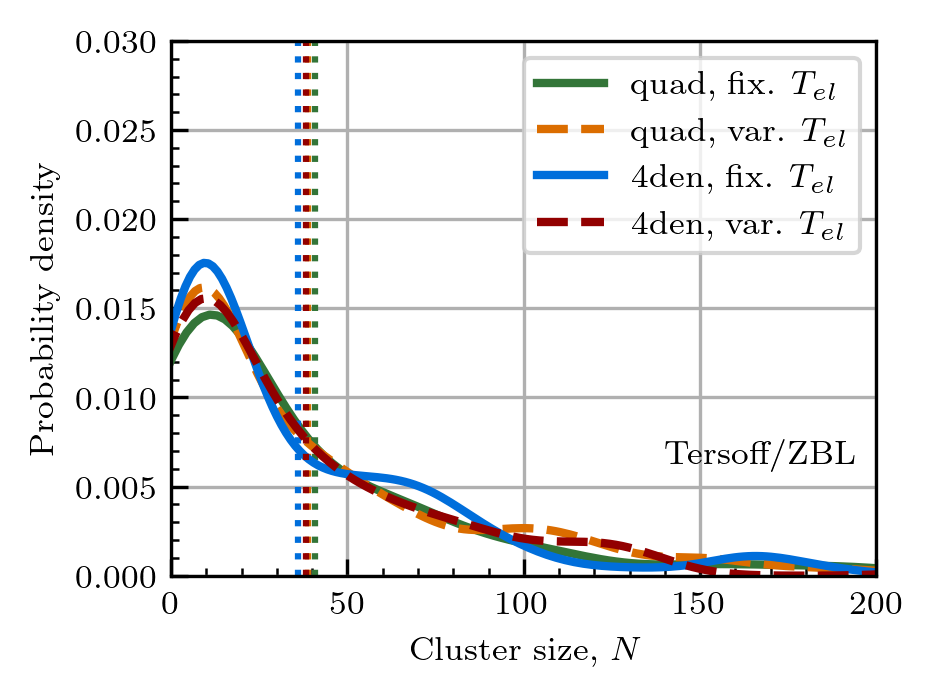}
		\caption{Distribution of cluster length along the first principal component and cluster size (number of defects per cluster).}
		\label{fig:probability-density}
	\end{figure*}

	\begin{figure*}[h!]
		\centering	
		\includegraphics[scale=1]{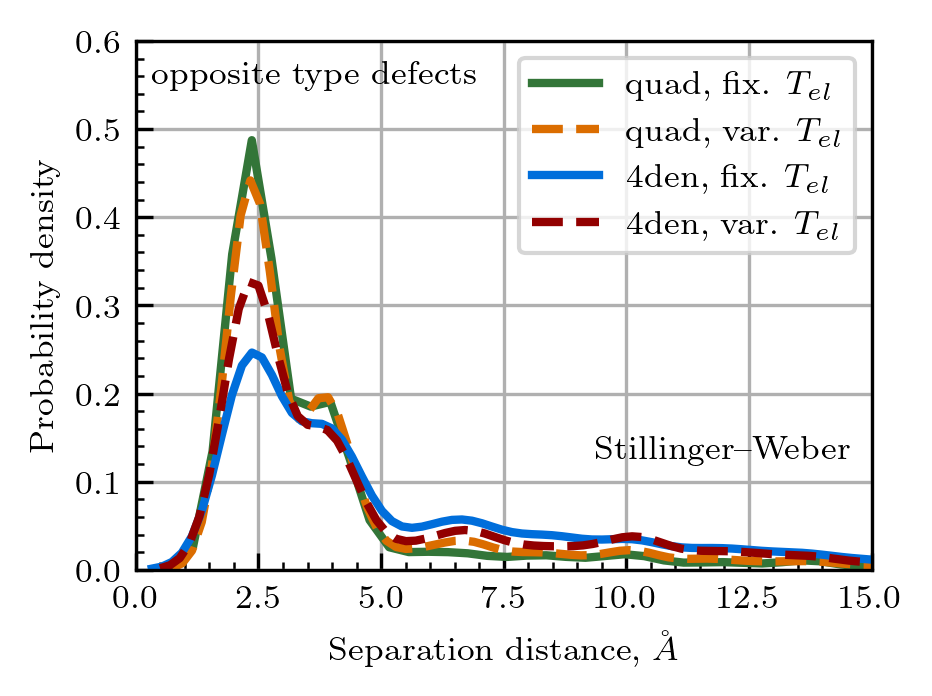}
		\includegraphics[scale=1]{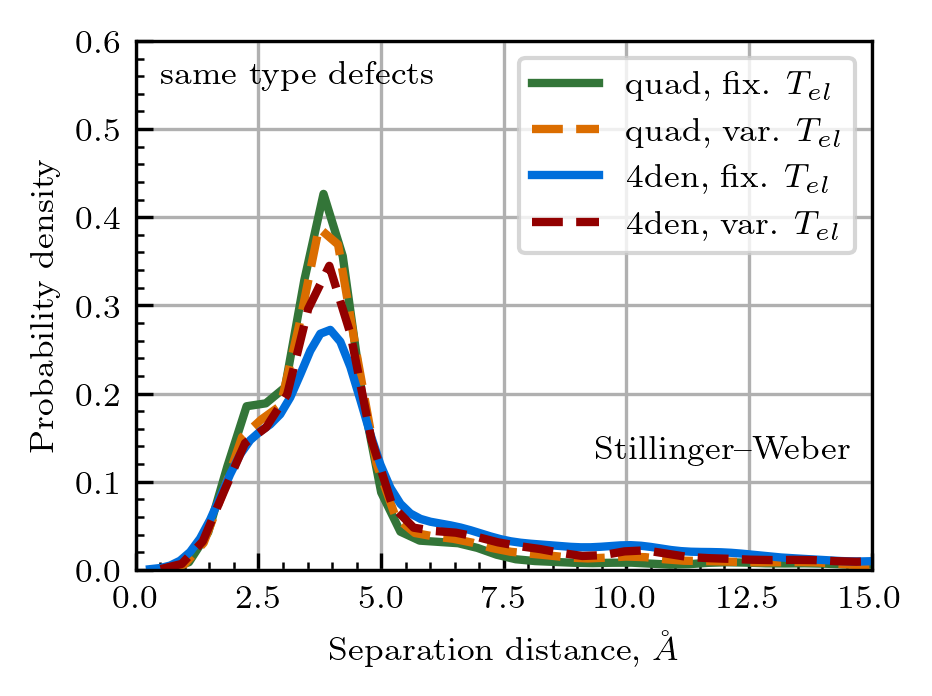}
		\includegraphics[scale=1]{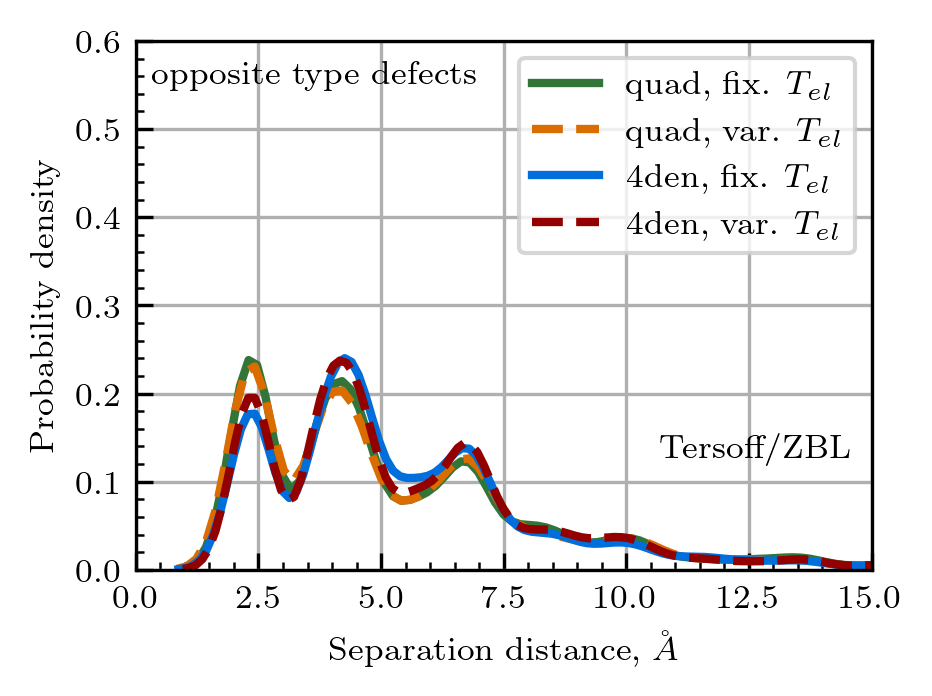}
		\includegraphics[scale=1]{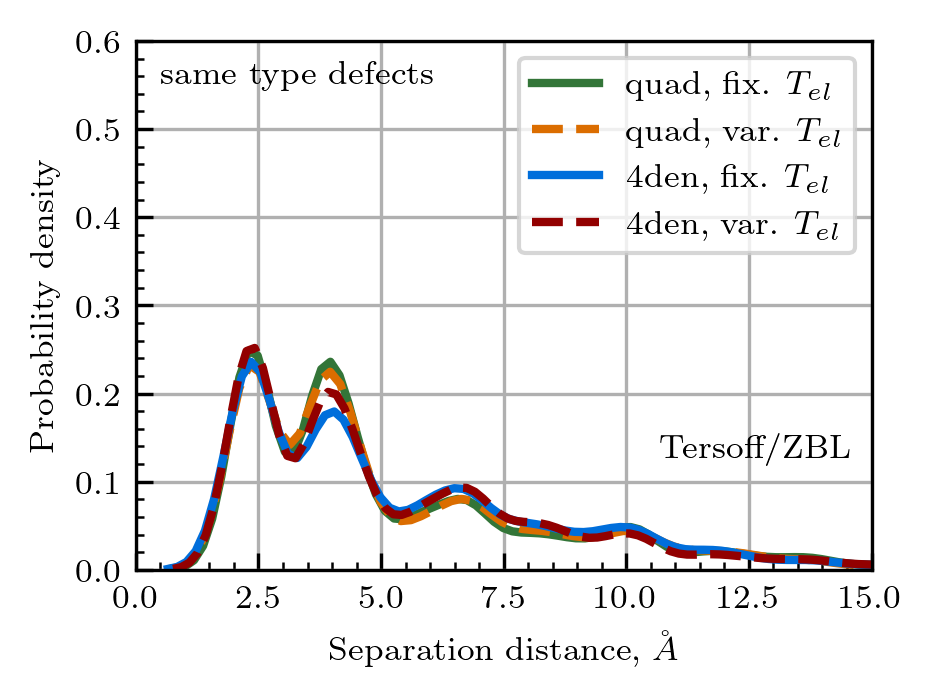	}
		\caption{Probability density function of the defect-defect separation distances. The left column shows separations between defects of the opposite type, while the right column shows separations between defects of the same type.}
		\label{fig:nn}
	\end{figure*}

	Correct predictions of cluster sizes in the primary damage are crucial for modelling long-term defect evolution, as clusters behave differently from isolated point defects, leading to qualitative and quantitative differences in long time damage accumulation and annealing.
	
	The observed differences in energy dissipation between the models investigated here directly influence defect clustering as shown in Fig.~\ref{fig:clustering}.
	The four-density coupling function consistently predicts less clustering compared to the quadratic coupling. We observe a stronger effect in simulations with the SW potential, due to the differences in cascade morphology predicted with the different potentials.
	
	We also compare the recombination efficiency, calculated as:
	\begin{equation}
		R = 1 - \frac{N_{surv}}{N_{max}}
		\label{eq:recrystal}
	\end{equation}
	where $N_{surv}$ is the number of surviving coordination defects in a cascade and $N_{max}$ is the maximum number during peak disorder in the same cascade. A recombination efficiency of 1 indicates complete recombination, with no surviving defects.
	
	The recombination efficiency is presented in Fig.~\ref{fig:recysltal}. For the T3/ZBL potential, the four-density coupling reduces the recombination efficiency relative to the quadratic function. In contrast, for the SW potential, the four-density coupling function significantly increases the recombination efficiency compared to the quadratic function. This behavior arises from differences in cascade morphology as modeled by the interatomic potentials, with differences in how each interatomic potential stabilizes or penalizes specific configurations.

	In Fig.~\ref{fig:dimer-poteng}, we present the potential energy of a dimer for the SW and T3/ZBL potentials. For comparison, we also present the potential energy of a dimer calculated using only the ZBL potential and results obtained from DFT calculations \cite{Saha2026}. The T3/ZBL potential is combined with the ZBL potential, thus, it follows ZBL behavior for distances less than 1~\AA. The SW potential has stronger repulsion compared to T3/ZBL, which translates into a smaller penetration depth of recoils.
	
	Consequently, the SW potential produces compact, high-density cascades, whereas cascades in the T3/ZBL potential are more spatially extended.
	These differences translate into cluster geometry. Fig.~\ref{fig:probability-density} shows that T3/ZBL tends to produce more elongated clusters, while SW clusters are more compact.

	At the same time, the T3/ZBL potential gives relatively low energies for coordination defects \cite{Nordlund1998}, which allows damage formed in the ballistic phase of the cascade to freeze in with almost no recombination.	
	The SW potential, on the other hand, is fitted mostly to the tetrahedral configuration and penalizes non-tetrahedral bonding types \cite{Nordlund1998}.

	These properties translate into differences in spatial correlations between defects. 	
	Fig.~\ref{fig:nn} shows the probability density functions of the defect-defect separation distance for defects of the same type and of opposite types. For the SW potential, defects of opposite types are generally located closer to each other than defects of the same type, which increases the probability of defect encounters and recombination compared to the T3/ZBL potential. In the T3/ZBL potential, the probability density distributions are similar for both same-type and opposite-type defect pairs. Interestingly, the T3/ZBL potential exhibits three distinct peaks in the separation-distance distribution, indicating preferred defect positions and a more complex internal cluster structure. In contrast, the SW potential produces a more homogeneous distribution, resulting in damage that is more amorphous-like.
	
	For the SW potential, this results in the four-density coupling function reducing the number of defects within clusters while the cluster length remains approximately unchanged (see Fig.~\ref{fig:probability-density}), indicating that SW clusters form amorphous pockets that do not fully recrystallize, despite containing fewer defects. In contrast, the more complex cluster structure produced by the T3/ZBL potential makes damage recombination more difficult.

	\section{Conclusions}
    \label{sec:conclusion}
	
	In this work, we investigated the role of electronic energy losses in the formation of primary radiation damage in Si using two interatomic potentials (SW and T3/ZBL) and two different ion-electron coupling functions for two-temperature molecular dynamics simulations.
	
	The choice of the coupling function is found to have an impact on defect production, clustering, and recombination. However, in some aspects the magnitude and even direction of these effects depend on the interatomic potential, indicating that both the electronic coupling model and the atomic interaction model must be considered together for accurate predictions.

	The UTTM-MD simulations with the four-density coupling function systematically predict a lower number of surviving defects and reduced clustering.
	The observed differences arise from significantly enhanced electronic energy losses for the four-density coupling function, which leads to reduced energy available for defect formation.
	
	Despite both coupling functions being fitted to rt-TDDFT data, they produce significantly different predictions under cascade conditions. We find better agreement of the four-density coupling with energy losses predicted using SRIM stopping powers in the ballistic regime, suggesting a more physically consistent description of electronic energy dissipation under high-energy collision conditions.

	\section{Acknowledgements}
	
	This work was supported by the Emil Aaltonen Foundation and by the European Union (ERC-2022-STG, project MUST, No. 101077454). Views and opinions expressed herein are those of the authors only and do not necessarily reflect those of the European Union or the European Research Council Executive Agency. Neither the European Union nor the granting authority can be held responsible for them. Calculations were performed using computer resources within the Aalto University School of Science ``Science-IT'' project.

	\section{Declaration of generative AI and AI-assisted technologies in the manuscript preparation process}
	
	During the preparation of this work the authors used ChatGPT in order to improve language and readability. After using this tool/service, the authors reviewed and edited the content as needed and take full responsibility for the content of the publication.
    
    \bibliographystyle{unsrt}
	\bibliography{/home/nadia/yandex.disk/bib-database/radiation-damage-finland,/home/nadia/yandex.disk/bib-database/silicon-radiation-experiments}

\end{document}